# MEASURING THE TECHNOLOGICAL PEDAGOGICAL CONTENT KNOWLEDGE (TPACK) OF IN-SERVICE TEACHERS OF COMPUTER SCIENCE WHO TEACH ALGORITHMS AND PROGRAMMING IN UPPER SECONDARY EDUCATION

Spyros Doukakis
Dept. of Primary Education
University of the Aegean

Alexandra Psaltidou and Athena Stavraki
The American College of Greece, Pierce College

Nikos Adamopoulos
ICT Center of Primary and Secondary Education of Ileia

Panagiotis Tsiotakis
ICT Center of Primary and Secondary Education of Korinthia

Stathis Stergou
1st Lyceum of Rhodes
Greece

**Abstract**
Based on the Technological Pedagogical and Content Knowledge (TPACK) framework (Mishra & Koehler, 2006) and the Schmidt et al. (2009) instrument which explore TPACK, this study examines a national sample of 1032 secondary teachers of computer science and measures their knowledge with respect to technology, pedagogy, content knowledge and the combination of each of these areas. Findings indicate that content knowledge and technology knowledge rating are high (average 4.38 and 4.16 respectively) and it seems that secondary teachers are less confident with their pedagogical content knowledge and their technological content knowledge (average 3.51 and 3.68 respectively).

## Introduction

The use of technological tools as educational aids has been playing an increasingly important role in all subjects in primary and secondary education in recent years (Polly, Mims, Shepherdd, & Inan, in press). According to Graham et al. (2009), educators have realized that a basic knowledge of how to use technological tools is not enough; what they really need is the ability to use these tools effectively in order to facilitate the learning processes of their students. Thus, research interest has focused on ways to incorporate and integrate technological tools in teaching. Recently, Koehler and Mishra (2005) reported that "we view teacher knowledge about technology as important, but not separate and unrelated from contexts of teaching." Mishra and Koehler (2006) have built upon Shulman's (1987) work describing Pedagogical Content Knowledge (PCK) and proposed a framework combining three important aspects of teacher knowledge: Pedagogical Knowledge, Content Knowledge and Technological Knowledge called the Technological Pedagogical and Content Knowledge (TPACK) framework. There is a wealth of research related to TPACK in the domain of maths, arts, social studies, sciences and English as a foreign language (e.g. Cavin, 2007; Guzey & Roehrig, 2009; Keating & Evans, 2001; Koehler & Mishra 2005; Niess, 2005; Richardson, 2009; Van Olphen, 2008).

For many years in many different countries computing has been included in the curriculum as a distinct discipline in secondary education. Computing focuses on how computers work (hardware) and how to program them (programming and software development), whereas ICT (Information and Computer Technology) is focused on how to use computers. Teachers of Computer Science (CS) in secondary education belong to a special category with a broad knowledge of computing and ICT based on their academic studies. The ACM K–12 Education Task Force Report (2003) draws attention to the need for appropriate Computer Science teacher training programs and notes "teachers must acquire both a mastery of the subject matter and the pedagogical skills that will allow them to present the material to students at appropriate levels" (p. 18). However, in the TPACK framework, it is important for teachers of CS to be able to integrate in their teaching practice technological tools, in such a way that these tools will not only form the subject matter but also a means to facilitate the learning process.

In this study using quantitative methods a) we measure the level of integration of technological tools and b) we explore how technological tools for teaching algorithmic and programming disciplines are integrated in the teaching practices.

## Teaching Algorithms and Programming in Upper Secondary Education in Greece

In Greece, the teaching of Computing and ICT in secondary education is conducted by teachers holding an undergraduate degree in Computer Science, Computer Engineering or Applied Informatics. Secondary Education in Greece is divided into two cycles: compulsory lower secondary and non-compulsory Upper Secondary Education. Compulsory lower secondary education is provided in Gymnasio, while non-compulsory upper secondary education is provided in one of two types of schools: the General Lykeio (GL) and Vocational Lykeio (EPAL). Parallel to these full time day schools are evening secondary schools. The duration of Gymnasio, both day and evening, is three years. The duration of studies in a General Lykeio (GL) is three years, unless it operates as an evening school, in which case it is four years.

Computing and ICT courses are mandatory during Gymnasio years and aim to develop students' skills in the use of ICT (operating systems, word processing, spreadsheets, image processing etc). In the third year of Gymnasio students acquire fundamental algorithmic and programming skills within a LOGO based environment.

Holders of a school-leaving certificate from a Gymnasio may register in a General Lykeio or in a Vocational Lykeio, without entry exams or other limitations (Eurydice Network, 2010).The curriculum at Vocational Lykeio includes general knowledge subjects, technical — vocational subjects and workshop exercises. With respect to Computing and ICT, Vocational Lykeio offer an Information Science sector with the following specialization: "Computer systems & networks electronic experts" and "System, Applications and Computer Networks' Support."

The 1st grade of General Lykeio operates as an orientation year with a general knowledge programme. The 2nd grade offers three curricular directions or pathways: Theoretical, Scientific and Technological. In the 3rd grade General Lykeio again has directions/pathways but the Technological Direction provides two courses: i) the Technology and Production course and ii) the Information Science and Services course (Eurydice Network, 2010).

In the third grade, students who follow the technological direction of the Information Science and Services course will take a course which involves the development of algorithms and programming (Applications Development in a Programming Environment (ADPE)). This course has been taught for ten years. It focuses on the algorithmic approach and on the development of problem-solving skills in a programming environment, rather than on programming techniques and the learning of a specific programming language. This subject is also assigned to CS teachers.

The overall aim of ADPE is to develop analytical and synthetic thinking, acquire methodological skills and be able to solve simple problems within a programming environment. Concepts such as learning a particular programming environment or examining the detailed structure and the syntax rules of a programming language do not comply with the aim of this subject. In other words, ADPE has not been designed to create programmers, and for this reason it not designed to teach sophisticated programming techniques; it focuses on approaches and techniques of problem solving with emphasis on structured thinking (Vakali et al., 1999). Many basic algorithmic and programming concepts as conditions, expressions and logical reasoning, are fundamentals of general knowledge and skills to be acquired in general education; most of these concepts are not presented in other disciplines (Dagdilelis, Satratzemi, & Evangelidis, 2004; Politis & Komis, 1999; Voyiatzaki, Christakoudis, Margaritis, & Avouris, 2004).

The curriculum states that this subject must be taught (at least partially) in a computer lab. The Pedagogical Institute has certified specific Educational Software to support the lab work in this course. The Educational Software has been designed to support teaching, to complement the subject's needs and IT use and to help students consolidate the material. The certified software includes an activity space, a flow chart developer and a programming environment in accordance with the textbook. In addition, two more educational software packages have been developed by educators and are already in use. During the lab hour, teachers can use the technological tools to facilitate the learning process. Furthermore, students sit for examinations in the subject which are carried out in a national level at the end of the school year. The grade acquired in this examination is part of the consideration used in selecting students for admission in higher education programmes.

In this complex framework of ADPE, where examination pressure and requirements coexist with the use of technological tools for teaching algorithmic concepts, it was considered essential to investigate the TPACK of secondary teachers of Computer Science who teach the subject.

## PCK and TPACK in Teaching Algorithms and Programming

PCK is a framework that views knowledge of content (e.g. maths, computer science, arts, science, etc.) in conjunction with knowledge of the pedagogy (how to teach), giving insights into educational matters relative to the learning and teaching of a topic. Teachers with good PCK are teachers who can transform their knowledge of the subject and make it accessible to their learners. PCK also includes an understanding of difficulties that may arise in learning special topics (Ragonis & Hazzan, 2008).

The PCK framework adapted to teaching programming includes the following components (Jimoyiannis, 2005):

- Content knowledge, for subject matter.

- Knowledge and perceptions of the goals, objectives, means and strategies of teaching programming at every level (knowledge of the curriculum).

- Knowledge of methods of understanding, perception, difficulties and misunderstandings encountered by students in specific units of programming theory.
- Knowledge of appropriate models of knowledge, available educational means and effective teaching strategies for each unit.

- Knowledge and perceptions about how to evaluate the scientific literature on programming and teaching approaches for programming.

Recently, research in educational technology suggests the need for "Technological Pedagogical and Content Knowledge" (TPACK), which is based on Shulman's (1987) idea of "Pedagogical Content Knowledge", so as to incorporate technology in pedagogy (Angeli & Valanides, 2009; Mishra & Koehler, 2006). This interconnectedness among content, pedagogy and technology has important effects on learning as well as on professional development. Mishra and Koehler (2006) suggest ". . .a curricular system that would honour the complex, multi-dimensional relationships by treating all three components in an epistemologically and conceptually integrated manner" (p. 1020). They have proposed a model that suggests three unitary components of knowledge (Content, Pedagogy and Technology), three dyadic components of knowledge (Pedagogical Content, Technological Content, Technological Pedagogical) and one overarching triad (Technological Pedagogical Content Knowledge).

Figure 1: Framework of technological pedagogical and content knowledge

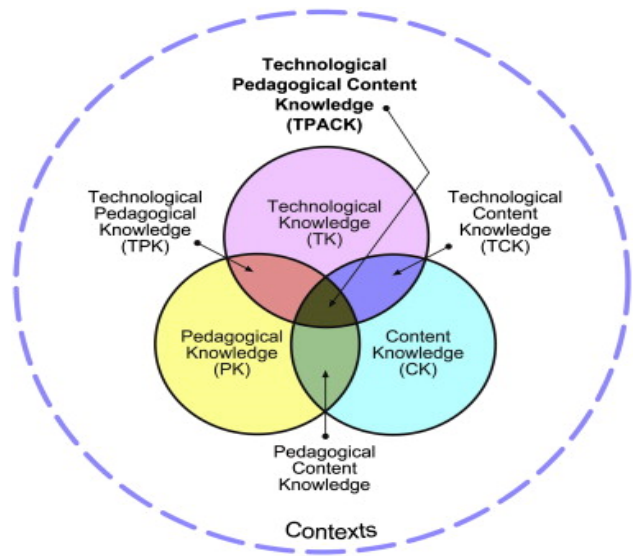

(Source: Koehler & Mishra, 2005)

According to Polly et al. (in press),

> In order to effectively integrate technology into their classroom, teachers must be knowledgeable about the relationships between technology and content — how technology can be used to support the learning of specific content, technology and pedagogy- how specific pedagogies best support the use of technology, and content and pedagogy- how specific pedagogies facilitate learning of specific content. Further, the center circle recognizes that teachers must possess knowledge about the intersection between technology, pedagogy as well as content that they are teaching.

The TPACK of in-service and pre-service teachers has been measured in both qualitative and quantitative studies (Archambault & Crippen, 2009; Doering, Scharber, Miller, & Velesianos, 2009; Schmidt et al., 2009). However, no study had been conducted for the TPACK of teachers of Computer Science. Therefore, the purpose of this paper is to explore TPACK of teachers of Computer Science in secondary education.

## Research Context and Participants

We conducted a quantitative survey using a questionnaire with closed-ended questions. The questionnaire was available only through Internet browsers and participants answered electronically on-line (http://www.kwiksurveys.com/online-survey.php?surveyID=LONJN_79a061ec).

The survey used a sample of 1032 teachers who had taught ADPE; 635 of the participants completed the entire questionnaire (62%). The questionnaire consisted of 29 questions about TPACK and is based on the survey instrument developed by Schmidt et al. (2009). All questions are related to the three key domains as described by the TPACK framework (technology, pedagogy, content and the combination of each of these areas). The 29 questions in the questionnaire are divided into questions about CK (4 questions), TK (4 questions), PK (4 questions), PCK (2 questions), TCK (5 questions), TPK (7 questions) and TPACK (3 questions). The response to each question is scored using a Likert like scale where 1 is strongly disagree, and 5 is strongly agree. For each subscale (CK, TK, PK, PCK, TCK, TPK, TPACK) the participant's responses are averaged. In addition, the questionnaire included 10 questions which provided demographic data.

In addition to the survey instrument participants filled out a questionnaire with 196 questions, all concerning the specific subject (ADPE). Some of these answers were evaluated and used for this study. Participants completed the questionnaire between November 2009 and January 2010; in January teachers are half way through the curriculum, having covered the appropriate material for all basic algorithmic components (sequential structure, conditional structure and loops). During this period, educators have a significant workload.

There are 1712 CS teachers in upper secondary education in Greece (National Statistical Service of Greece, 2009). The respondents come from the 13 regions of the country. The sample is representative of the population of educators of Computer Science by region.

## Results

The majority of participants (74%) were teaching the lesson in the current school year, 14% taught ADPE the previous year and the remainder (12%) the year before that. Of the 635 teachers who completed the survey, 214 (33.6%) were female and 421 (66.4%) were male. Approximately 24% of the respondents have been teaching the subject for three years (1–3 years of service), 30% for 4–6 years, 27% for 7–9 years, while 19% have been teaching the course from the first year that it was introduced in the technological direction. The majority of teachers (61%) have at most an undergraduate degree, while 35% have at most a postgraduate degree and the remainder have PhDs. The majority of the participants (90%) teach the subject in public or private schools and 10% tutor privately or teach in private tutorial centres.

The average mean for all items was 4.05. The range of response was 4, with a minimum response of 1, a maximum response of 5 and a standard deviation of .805. The respondents answered all the questions; the mean and standard deviation are reported for each subscale in Table 1.

Table 1: Summary of descriptive statistics for subscales of TPACK components

| Subscale | Number of Items | Number of Responses | Mean | Std. Deviation |
|---|---|---|---|---|
| CK | 4 | 635 | 4.38 | .488 |
| PK | 4 | 635 | 4.12 | .533 |
| TK | 4 | 635 | 4.16 | .552 |
| PCK | 2 | 635 | 3.51 | .692 |
| TCK | 5 | 635 | 3.68 | .802 |
| TPK | 7 | 635 | 4.18 | .511 |
| TPACK | 3 | 635 | 4.03 | .657 |

In addition to descriptive statistics measuring Computer Science Teachers' TPACK, correlations among each of the subscales using Pearson product-moment were examined (Table 2). With respect to correlations between subscales, coefficients varied from .235 (TCK and PCK) to .746 (PK and TK).

Table 2: Correlations among TPACK subscales

|  | TK | CK | PK | PCK | TCK | TPK | TPACK |
|---|---|---|---|---|---|---|---|
| TK | - |  |  |  |  |  |  |
| CK | .484 | - |  |  |  |  |  |
| PK | .746 | .547 | - |  |  |  |  |
| PCK | .428 | .312 | .378 | - |  |  |  |
| TCK | .318 | .316 | .282 | .235 | - |  |  |
| TPK | .494 | .403 | .456 | .388 | .322 | - |  |
| TPACK | .507 | .414 | .465 | .331 | .352 | .715 | - |
| Correlation is significant at the 0.01 level (2-tailed). | | | | | | | |

## Discussion

The teachers of Computer Science who participated in this survey rated knowledge of the subject matter (4.38) higher than that of the other cognitive subscales. This implies that teachers have very high Content Knowledge, such as knowledge of algorithmic concepts presented in the subject, theories, the general framework of the discipline and practical approaches used in order for students to acquire knowledge (Shulman, 1987).

According to their responses, teachers seem to have very high (4.18) Technological Pedagogical Knowledge (TPK, 4.18). This shows that educators have realized that teaching and learning are reformed when using specific technological tools. This knowledge includes awareness of tools' limitations and capabilities in designing pedagogical strategies. The TPK is likely to appear stronger because all available technological tools are designed to fulfil educational aims of the subject (Koehler & Mishra, 2008).

Technological Knowledge (4.16) is also very high. According to Koehler and Mishra (2008) Technological Knowledge is associated with the ability to use technological tools but also the knowledge behind this technology. This intersection enables teachers to apply technological knowledge effectively for the benefit of student learning and to be open to forthcoming changes.

The Pedagogical Knowledge rating (4.12) is very close to that of Technological Knowledge. The high average implies that CS teachers have deep knowledge of the educative process and methodology of teaching and learning and thus can meet the aim of the subject. In this way, they are able to contribute to student learning, manage the classroom, create course outlines and educational scenarios and conduct student assessment.

The average scores are lower (3.68 and 3.51 respectively) for TCK and PCK. For TCK, it seems that teachers rate themselves with a lower level in the understanding of how technology and subject matter both aid, and limit each other. Therefore, teachers seem to need assistance in order to comprehend how technology use affects the subject matter. Even if the high score in TPK shows that teachers seem to be aware of the effect of technology on teaching and learning, they can not really identify the manner in which technology does this. It is necessary to distinguish which technological tool is the most appropriate to support a specific cognitive area of algorithmic development and how the content leads and/or changes technology.

The last dyadic component (PCK) shows that, even if teachers of CS have both pedagogical knowledge and deep knowledge of their subject matter, they seem to be less confident in transforming and applying effectively their Content Knowledge in their teaching process. According to Shulman (1987) this transformation occurs when the teacher is able to interpret the content of the subject, find alternative ways to present it and adapt the educational material to students' perception and prior knowledge. For the development of PCK, teachers should be perceptive in recognising students' common misconceptions and the methods by which these misconceptions can be deconstructed. Teachers should be able to draw from different cognitive areas and be flexible in trying out alternative approaches.

Finally, the high score in TPACK (4.03) shows that teachers are aware of the intersection between content, pedagogy and technology. The TPACK according to Koehler and Mishra (2008), forms the foundation of effective teaching with technology use and requires an understanding "of the representation of concepts using technologies; pedagogical techniques that use technologies in constructive ways to teach content; knowledge of what makes concepts difficult or easy to learn and how technology can help redress some of the problems that students face; knowledge of students' prior knowledge and theories of epistemology; and knowledge of how technologies can be used to build on existing knowledge and to develop new epistemologies or strengthen old ones." Thus, it appears that teachers enhance teaching with a unique combination, a dynamic equilibrium, between the three teaching components (pedagogy, content and technology).

Despite the fact that CS teachers who teach this subject claim to posses the above qualities, it seems that only 62% use technological tools and computer laboratory, while 38% teach the subject exclusively in the classroom. Among those who do not use the computer laboratory, 40% do not use computers at all, not even in the presentation of material to students. In relevant questions about why they do not use the computer labs, the majority attribute it to time pressure (68%). ADPE is scheduled to be taught for two periods (40–45 minute) per week, while 95% of teachers consider at least 3 periods as the minimum necessary. Additionally, teachers state that this extra hour is necessary in order to use the computer labs (70%). A second reason why teachers do not use the laboratory is the number of students per class; classes have many students and it is often difficult to accommodate them in the laboratory (15%). Other reasons are the lack of well equipped laboratories (8%), lack of appropriate educational materials and scenarios (7%) and the fact that the national examination is conducted on paper rather than on a computer (2%).

For the 62% of teachers who use technological tools and the laboratory, 65% of them consider that conducting sessions in the laboratory reduces the time needed to cover the curriculum. However, they use technological tools mostly to present algorithmic issues (41%) and less for students' practice with the available tools and relevant training scenarios (31%).

This is also reflected in questions concerning the type of training they consider is necessary. A percentage of 37% of the respondents needs training in educational software and 70% in methods to integrate educational software in their teaching practice. Finally, 43% need training in teaching methods of algorithmic structures.

## Conclusion

Teachers of Computer Science belong to a special group with a highly developed knowledge of technology. The aim of the survey was to compile data for this group according to the TPACK framework. According to the results in the seven subscales, teachers state that their knowledge is between the values 4.38 (CK) and 3.51 (PCK).

Data collected can help set guidelines for training programs for future CS teachers. Moreover, according to the results, teachers try to find a teaching framework beyond the traditional classroom that incorporates the use of the computer lab. This idea is consistent with the subject ADPE. It seems that even if teachers of Computer Science have more experience and knowledge in using computers compared to teachers from other disciplines, they require training on methods to integrate technological tools into their teaching. On the other hand, special attention should be given towards the development of appropriate educational scenarios and examples from every day life that can improve both students' learning and teachers' work.